\documentclass[final,1p,times,authoryear]{elsarticle}
\usepackage{amsmath}
\usepackage{amssymb}
\usepackage{tikz}
\usepackage{url}

\let\set\mathbb
\def\<#1>{\langle#1\rangle}

\def\O{\operatorname{O}}
\def\sgn{\operatorname{sgn}}
\def\gl{\operatorname{GL}}
\def\rank{\operatorname{rank}}
\def\elldown{^{\text{\tiny (}\iota\text{\tiny )}}}
\def\ell{\iota}
\def\neg{\bar}

\def\eatspace#1{#1}
\def\step#1#2{\par\kern1pt\dimen44=#2em\advance\dimen44 1.67em\hangindent\dimen44\hangafter=1\noindent\rlap{\small#1}\kern\dimen44\relax\eatspace}

\begin{document}

\begin{frontmatter}

\title{New ways to multiply $3\times 3$-matrices\footnote{%
    M.J.H. Heule was supported by NSF under grant CCF-1813993.
    M. Kauers was supported by the Austrian FWF grants P31571-N32 and F5004.
    M. Seidl was supported by the Austrian FWF grant NFN S11408-N23 and  the LIT AI Lab funded by the State of Upper Austria.
    }
}

 \author{Marijn J.H. Heule}
 \address{Department of Computer Science, The University of Texas, Austin TX, USA}
 \ead{marijn@cs.utexas.edu}

 \author{Manuel Kauers}
 \address{Institute for Algebra, J. Kepler University Linz, Austria}
 \ead{manuel.kauers@jku.at}

 \author{Martina Seidl}
 \address{Institute for Formal Models and Verification, J. Kepler University Linz, Austria}
 \ead{martina.seidl@jku.at}

 \begin{abstract}
   It is known since the 1970s that no more than 23 multiplications are required
   for computing the product of two $3\times 3$-matrices.
   It is not known whether this can also be done with fewer multiplications.
   However, there are several mutually inequivalent ways of doing the job with 23 multiplications.
   In this article, we extend this list considerably by providing more than $13\,000$ new and mutually
   inequivalent schemes for multiplying $3\times 3$-matrices using 23 multiplications.
   Moreover, we show that the set of all these schemes is a manifold of dimension at least~17.
 \end{abstract}

 
\end{frontmatter}

 \section{Introduction}

 The classical algorithm for multiplying two $n\times n$ matrices performs
 $2n^3-n^2$ additions and multiplications. Strassen's algorithm~\citep{strassen1969gaussian}
 does the job with only $\O(n^{\log_27})$ additions and multiplications, by
 recursively applying a certain scheme for computing the product of two $2\times
 2$-matrices with only 7 instead of the usual 8 multiplications. The discovery
 of Strassen's algorithm has initiated substantial work during the past 50 years
 on finding the smallest exponent $\omega$ such that matrix multiplication costs
 $\O(n^\omega)$ operations in the coefficient ring.
 The current record is $\omega\leq 2.3728639$ and was obtained by~\cite{LeGall:2014:PTF:2608628.2608664}.
 It improves the previous record of~\cite{Williams:2012:MMF:2213977.2214056} by just $3\cdot10^{-7}$.
 Extensive background in this direction is available in text books~\citep{buergisser-book,landsberg2017geometry} and
 survey articles~\citep{blaeser-survey2013,pan-survey2018}. 
 Contrary to wide-spread belief, Strassen's algorithm is not only efficient in theory but also in practice. 
 Special purpose software for exact linear algebra, such as the FFLAS and FFPACK packages~\citep{DGP:2008},
 have been using it since long, and there are also reports that its performance in a numerical context
 is not as bad as its reputation~\citep{Huang:2016:SAR:3014904.3014983}.
 
 Besides the quest for the smallest exponent, which only concerns the asymptotic
 complexity for asymptotically large~$n$, it is also interesting to know how many
 multiplications are needed for a specific (small) $n$ to compute the product of two $n\times
 n$-matrices. Thanks to Strassen, we know that the answer is at most 7 for
 $n=2$, and it can be shown~\citep{winograd1971multiplication} that there is no way to do it with 6
 multiplications. It can further be shown that, in a certain sense, Strassen's scheme
 is the only way of doing it with 7 multiplications~\citep{de1978varieties}. 

 Already for $n=3$, the situation is not completely understood. \cite{laderman1976noncommutative}
 showed that 23 multiplications suffice, and \cite{blaser2003complexity} showed that at least 19
 multiplications are needed.
 For larger sizes as well as rectangular matrices, many people have been searching
 for new schemes using fewer and fewer coefficient multiplications.
 For $n=4$, the best we know is to apply Strassen's scheme recursively, which requires~49 multiplications.
 For $n=5$, the record of 100 multiplications was held~\cite{MAKAROV1987205} for 30 years until it was
 improved to 99 by~\cite{DBLP:journals/corr/Sedoglavic17aa}. 
 For $n=6$, there is a recent scheme by~\cite{smirnov2013bilinear} which needs only 160 multiplications.
 For $n=7$, \cite{DBLP:journals/corr/abs-1712-07935} found a way to compute the product with 250 multiplications. 
 For larger sizes and rectangular matrices, see the extensive tables compiled by~\cite{smirnov2013bilinear,smirnovseveral}
 and~\cite{fastmm}. 
 Many of the schemes for larger matrix sizes are obtained by combining
 multiplication schemes for smaller matrices~\citep{DBLP:journals/tcs/DrevetIS11}. 
 
 Although nobody knows whether there is a scheme using only 22 multiplications for~$n=3$ (in an exact and
 non-commutative setting), 23 multiplications can be achieved in many different
 ways. \cite{DBLP:journals/siamcomp/JohnsonM86} have in fact found infinitely many ways.
 They presented a family of schemes involving three free parameters.
 However, their families involve fractional coefficients and therefore
 do not apply to arbitrary coefficient rings~$K$. Many others have reported isolated
 schemes with fractional or approximate coefficients. Such schemes can be 
 constructed for example by numerically solving a certain optimization problem,
 or by genetic algorithms. In Laderman's multiplication scheme, all coefficients are
 $+1$,~$-1$, or~$0$, which has the nice feature that it works for any
 coefficient ring. As far as we know, there are so far only three other schemes
 with this additional property, they are due to \cite{smirnov2013bilinear}, \cite{oh2013inequivalence}, and
 \cite{DBLP:journals/corr/abs-1108-2830}, respectively. We add more than 13\,000 new schemes to this list. 

 The isolated scheme presented by Courtois et al. was not found numerically but with the help of
 a SAT solver. SAT~\citep{DBLP:series/faia/2009-185} refers to the decision problem of propositional logic:
 given a Boolean formula in conjunctive normal form, is there an assignment of the Boolean variables
 such that the formula evaluates to true under this assignment? Although SAT is a prototypical example
 of an NP-complete problem, modern SAT solvers are able to solve very large instances. In addition to
 various industrial applications, they have recently also contributed to the solution of difficult
 mathematical problems, see \cite{DBLP:conf/sat/HeuleKM16} and~\cite{DBLP:conf/aaai/Heule18} for two
 examples. SAT solvers also play a central role in our approach. 
 As explained in Section~\ref{sec:searching}, we first use a SAT solver to find multiplication
 schemes for the coefficient ring~$\set Z_2$, starting from some known solutions. 
 In a second step, explained in Section~\ref{sec:sieving}, we discard solutions that
 are equivalent to solutions found earlier.
 Next, we simplify the new solutions (Sect.~\ref{sec:switching}), and use them as 
 starting points for a new round of searching.
 Altogether about 35 years of computation time were spent in several iterations of this process. 
 In the end, we lifted the solutions from $\set Z_2$ to arbitrary coefficient rings (Sect.~\ref{sec:signing}),
 and we extracted families with up to 17 free parameters from them (Sect.~\ref{sec:families}).
 Our 13,000 isolated schemes and our parameterized families are provided in various formats on our website~\citep{mmr}. 
 
 \section{The Brent Equations}\label{sec:brent}

 The general pattern of a matrix multiplication scheme consists of two sections.
 In the first section, several auxiliary quantities $M_1,M_2,\dots, M_m$ are computed,
 each of which is a product of a certain linear combination of the entries of the first matrix
 with a certain linear combination of the entries of the second matrix.
 In the second section, the entries of the resulting matrix are obtained as certain
 linear combinations of the auxiliary quantities $M_1,M_2,\dots, M_m$.

 For example, writing
 \[
 A = \begin{pmatrix}
   a_{1,1}&a_{1,2}\\
   a_{2,1}&a_{2,2}\\
 \end{pmatrix},\quad
 B = \begin{pmatrix}
   b_{1,1}&b_{1,2}\\
   b_{2,1}&b_{2,2}
 \end{pmatrix},\quad\text{and}\quad
 C = \begin{pmatrix}
   c_{1,1}&c_{1,2}\\
   c_{2,1}&c_{2,2}
 \end{pmatrix} := AB,   
 \]
 Strassen's multiplication scheme proceeds as follows:
 
 \hangindent=3em\hangafter=1\emph{First section.}\\
   $M_1 = (a_{1,1} + a_{2,2}) (b_{1,1} + b_{2,2})$\\
   $M_2 = (a_{2,1} + a_{2,2}) (b_{1,1})$\\
   $M_3 = (a_{1,1}) (b_{1,2} - b_{2,2})$\\
   $M_4 = (a_{2,2}) (b_{2,1} - b_{1,1})$\\
   $M_5 = (a_{1,1} + a_{1,2})(b_{2,2})$\\
   $M_6 = (a_{2,1} - a_{1,1}) (b_{1,1}+ b_{1,2})$\\
   $M_7 = (a_{1,2} - a_{2,2}) (b_{2,1} + b_{2,2})$

 \hangindent=3em\hangafter=1\emph{Second section.}\\
   $c_{1,1} = M_1 + M_4 - M_5 + M_7$\\
   $c_{1,2} = M_3 + M_5$\\
   $c_{2,1} = M_2 + M_4$\\
   $c_{2,2} = M_1 - M_2 + M_3 + M_6$.

 Observe that the number of multiplications is exactly the number of~$M$'s.
 Also observe that while it is not obvious how to construct such a scheme
 from scratch, checking that a given scheme is correct is an easy and
 straightforward calculation. For example, $c_{2,1}=M_2+M_4=(a_{2,1} + a_{2,2}) (b_{1,1})
 + (a_{2,2}) (b_{2,1} - b_{1,1}) = a_{2,1}b_{1,1} + a_{2,2}b_{2,1}$. 

 In order to search for a multiplication scheme for a prescribed shape of
 matrices (e.g., $3\times 3$) and a prescribed number of multiplications (e.g.,
 $23$), we can make an ansatz for the coefficients of the various linear
 combinations,
 \begin{alignat*}1
   M_1 &= (\alpha_{1,1}^{(1)}a_{1,1} + \alpha_{1,2}^{(1)}a_{1,2}+\cdots)(\beta_{1,1}^{(1)}b_{1,1} + \beta_{1,2}^{(1)}b_{1,2}+\cdots)\\
   M_2 &= (\alpha_{1,1}^{(2)}a_{1,1} + \alpha_{1,2}^{(2)}a_{1,2}+\cdots)(\beta_{1,1}^{(2)}b_{1,1} + \beta_{1,2}^{(2)}b_{1,2}+\cdots)\\
   &\vdots\\
   M_{23} &= (\alpha_{1,1}^{(23)}a_{1,1} + \alpha_{1,2}^{(23)}a_{1,2}+\cdots)(\beta_{1,1}^{(23)}b_{1,1} + \beta_{1,2}^{(23)}b_{1,2}+\cdots)\\
   c_{1,1} &= \gamma_{1,1}^{(1)}M_1 + \gamma_{1,1}^{(2)}M_2 + \cdots + \gamma_{1,1}^{(23)}M_{23}\\
   c_{1,2} &= \gamma_{1,2}^{(1)}M_1 + \gamma_{1,2}^{(2)}M_2 + \cdots + \gamma_{1,2}^{(23)}M_{23}\\
   &\vdots\\
   c_{3,3} &= \gamma_{3,3}^{(1)}M_1 + \gamma_{3,3}^{(2)}M_2 + \cdots + \gamma_{3,3}^{(23)}M_{23}
 \end{alignat*}
 and then compare coefficients such as to enforce $c_{i,j}=\sum_k
 a_{i,k}b_{k,j}$. Doing so leads to a system of polynomial equations for the
 undetermined coefficients $\alpha_{i_1,i_2}\elldown,\beta_{j_1,j_2}\elldown,
 \gamma_{k_1,k_2}\elldown$. The equations in this system are known as the Brent
 equations~\citep{brent1970algorithms}.  For $3\times 3$-matrices and 23 multiplications, the
 equations turn out to be
 \[
   \sum_{\ell=1}^{23} \alpha_{i_1,i_2}\elldown\beta_{j_1,j_2}\elldown\gamma_{k_1,k_2}\elldown = \delta_{i_2,j_1}\delta_{i_1,k_1}\delta_{j_2,k_2}
 \]
 for $i_1,i_2,j_1,j_2,k_1,k_2\in\{1,2,3\}$, i.e., there are 621 variables and
 729 cubic equations. The $\delta_{u,v}$ on the right refer to the
 Kronecker-delta, i.e., $\delta_{u,v}=1$ if $u=v$ and $\delta_{u,v}=0$
 otherwise.

 The equations become a bit more symmetric if we connect the matrices $A,B,C$
 through $C^\top=AB$ rather than $C=AB$. In the version with the transposition,
 which we shall use from now on, and which is also more common in the
 literature, the right hand side has to be replaced with
 $\delta_{i_2,j_1}\delta_{j_2,k_1}\delta_{k_2,i_1}$.

 In any case, the problem boils down to finding a solution of the Brent
 equations. In principle, this system could be solved using Gr\"obner
 bases~\citep{buchberger65,cox2013ideals,buchberger10}, but doing so would require an absurd amount of computation
 time. Some of the solutions reported in the literature have been found using
 numerical solvers~\citep{smirnov2013bilinear,oh2013inequivalence},
 and~\cite{laderman1976noncommutative} claims that his solution
 was found by solving the Brent equations by hand. He writes that he would explain in a later paper
 how exactly he did this, but apparently this later paper has never been written. Only recently,
 \cite{DBLP:journals/corr/Sedoglavic17} has
 given a convincing explanation of how Laderman's scheme can be derived from Strassen's scheme
 for $2\times 2$ matrices.
 \cite{DBLP:journals/corr/abs-1108-2830} found their solution using a SAT solver.
 We also start our search using SAT solvers.

 \section{SAT Encoding and Streamlining}\label{sec:searching}

 In order to encode the problem as a SAT
 problem, we view the Brent equations as equations for the finite field~$\set Z_2$, interpret 
 its elements as truth values, 
 its addition as exclusive {\sc or} ($\oplus$), 
 and its multiplication as
 conjunction ($\land$). These propositional formulas cannot be 
 directly be processed by most state-of-the-art SAT solvers, because they 
 require the formulas in conjunctive normal form (CNF). A formula is in 
 CNF if it is a conjunction of clauses, where a clause is a disjunction 
 ($\lor$) of 
 literals and a literal is a Boolean variable $x$ or the negation 
 of a Boolean 
 variable ($\neg x$). For avoiding an exponential blow-up when transforming an 
 arbitrary structured formula to CNF, auxiliary variables are introduced 
 that abbreviate certain subformulas. For every $i_1,i_2,j_1,j_2\in\{1,2,3\}$ and
 every $\ell=1,\dots,23$, we introduce a fresh variable $s\elldown_{i_1,i_2,j_1,j_2}$
 and impose the condition
 \[
 s_{i_1,i_2,j_1,j_2}\elldown\leftrightarrow (\alpha_{i_1,i_2}\elldown \land \beta_{j_1,j_2}\elldown),
 \]
 whose translation to CNF requires three clauses.
 Similarly, for every $i_1,i_2,j_1,j_2,k_1,k_2\in\{1,2,3\}$ and every
 $\ell=1,\dots,23$, we introduce a fresh variable $t_{i_1,i_2,j_1,j_2,k_1,k_2}\elldown$
 and impose the condition
 \[
 t_{i_1,i_2,j_1,j_2,k_1,k_2}\elldown\leftrightarrow (s_{i_1,i_2,j_1,j_2}\elldown \land \gamma_{k_1,k_2}\elldown),
 \]
 whose translation to CNF costs again three clauses.

 \def\even{\mathrm{even}}\def\odd{\mathrm{odd}}%
 For each fixed choice $i_1,i_2,j_1,j_2,k_1,k_2\in\{1,2,3\}$, there is a Brent equation
 which says that the number of $\ell$'s for which $t_{i_1,i_2,j_1,j_2,k_1,k_2}\elldown$
 is set to true should be even (if $\delta_{i_2,j_1}\delta_{i_1,k_1}\delta_{j_2,k_2}=0$)
 or that it should be odd (if $\delta_{i_2,j_1}\delta_{i_1,k_1}\delta_{j_2,k_2}=1$).
 It therefore remains to encode the condition that an even number (or an odd number) of a
 given set of $p$ variables should be true, i.e., we need to construct a formula
 $\even(x_1,\dots,x_p)$ which is true if and only if an even number among the
 variables $x_1,\dots,x_p$ is true. Such a formula can again be constructed using
 auxiliary variables. Note that $\even(x_1,\dots,x_p)$ is true if and only if
 $\even(x_1,\dots,x_i,y)\land\even(x_{i+1},\dots,x_p,y)$ is true, because this is the
 case if and only if both $\{x_1,\dots,x_i\}$ and $\{x_{i+1},\dots,x_p\}$ contain an
 even number of variables set to true (and then $y$ is set to false) or both sets contain
 an odd number of variables set to true (and then $y$ is set to true). Applying this principle
 recursively for $p = 23$ (the number of summands in each Brent equation), the problem can be broken down to chunks of size four:
 \begin{alignat*}5
   &\even(x_1,x_2,x_3,y_1)&&\land
   \even(x_4,x_5,x_6,y_2)&&\land
   \even(x_7,x_8,x_9,y_3)&&\land
   \even(x_{10},x_{11},x_{12},y_4)\\
   {}\land{}&\even(x_{13},x_{14},x_{15},y_5)&&\land
   \even(x_{16},x_{17},x_{18},y_6)&&\land
   \even(x_{19},x_{20},x_{21},y_7)&&\land
   \even(x_{22},x_{23},y_1,y_8)\\
   {}\land{}&\even(y_2,y_3,y_4,y_9)&&\land
   \even(y_5,y_6,y_7,y_{10})&&\land
   \even(y_8,y_9,y_{10},y_{11}).
 \end{alignat*}
 The small chunks can be encoded directly by observing that $\even(a,b,c,d)$ is equivalent
 to
 \begin{alignat*}1
 &(a \lor b \lor c \lor \neg d) \land
 (a \lor b \lor \neg c \lor  d) \land
 (a \lor \neg b \lor c \lor  d) \land
   (\neg a \lor b \lor c \lor  d) \land\\
 &(a \lor \neg b \lor \neg c \lor \neg d) \land
 (\neg a \lor \neg b \lor c \lor \neg d) \land
 (\neg a \lor b \lor \neg c \lor \neg d) \land
 (\neg a \lor \neg b \lor \neg c \lor d).
 \end{alignat*}
 For the cases where an odd number of the variables $x_1,\dots,x_{23}$ must be true, we can
 apply the encoding described above to $\even(\neg x_1,x_2,x_3,\dots,x_{23})$.

 The SAT problems obtained in this way are very hard. In order to make the problems more
 tractable, we added further constraints in order to simplify the search performed by the
 solver. This approach is known as streamlining~\citep{streamlining}. The following
 restrictions turned out to be successful:
 \begin{itemize}
 \item Instead of a faithful encoding of the sums in the Brent equations using the
   even predicate as described above, we also used a more restrictive sufficient condition 
   which instead of requiring an even number of arguments to be true enforces that zero or
   two arguments should be true. This predicate zero-or-two can be broken
   into at-most-two and not-exactly-one, which can be efficiently encoded as
   \begin{alignat*}1
     \textrm{not-exactly-one}(x_1,\dots,x_p) &= \bigwedge_{i=1}^p \Bigl(x_i \rightarrow \bigvee_{j\neq i} x_j\Bigr)\\
     \textrm{at-most-two}(x_1,\dots,x_p) &=
     (\neg x_1\lor\neg x_2\lor\neg x_3)\land(\neg x_1\lor\neg x_2\lor\neg x_4)\\
     &\quad\land(\neg x_1\lor\neg x_3\lor\neg x_4)\land(\neg x_2\lor\neg x_3\lor\neg x_4)\\
     &\quad\land(\neg x_1\lor y)\land(\neg x_2\lor y)\land(\neg x_1\lor\neg x_2\lor z)\\
     &\quad\land(\neg x_3\lor z)\land(\neg x_4\lor z)\land(\neg x_3\lor\neg x_4\lor y)\\
     &\quad\land\textrm{at-most-two}(y,z,x_5,\dots,x_p),
   \end{alignat*}
   where $y$ and $z$ are fresh variables. 
The first two lines of at-most-two assert that at most two variables 
of $x_1, x_2, x_3, x_4$ are true. 
If two or more of those variables are 
true then the new variables 
$y$ and $z$ have to be both true, if one variable is true, 
then either $y$ or $z$ has to be true, and if all four variables are false, 
then also $y$ and $z$ can be both false. Encoding this information 
in $y$ and $z$ allows to recursively apply at-most-two with two 
arguments less. 
A straightforward direct encoding as in the first two lines 
		 is used when $p\leq 4$.
 \item We selected a certain portion, say 50\%, of the variables $\alpha_{i_1,i_2}\elldown$, $\beta_{j_1,j_2}\elldown$, $\gamma_{k_1,k_2}\elldown$ and instantiate them with the values they have in one of the known solutions.
   The SAT solver then has to solve for the remaining variables. It turns out that in
   many cases, it does not just rediscover the known solution but finds a truly different one
   that only happens to have an overlap with the original solution.
 \item Another approach was to randomly set half of the terms $\alpha_{i_1,i_2}\elldown\beta_{j_1,j_2}\elldown\gamma_{k_1,k_2}\elldown$ with $i_2\neq j_1$ and $j_2\neq k_1$ and $k_2\neq i_1$ to zero. This strategy was
   motivated by the observation that in most of the known solutions, almost all these
   terms are zero.
 \item A third approach concerns the terms $\alpha_{i_1,i_2}\elldown\beta_{j_1,j_2}\elldown\gamma_{k_1,k_2}\elldown$ with
   $i_2=j_1$ and $j_2=k_1$ and $k_2=i_1$. Again motivated by the inspection of known solutions,
   we specified that for each $\ell$ either one or two such terms should be one.
   More precisely, we randomly chose a distribution of the 27 terms with $i_2=j_1$ and $j_2=k_1$ and $k_2=i_1$
   to the 23 summands of the scheme, with the condition that 19 summands should contain one term each
   and the remaining four summands should contain two terms each.
 \end{itemize}
 Each of the latter three approaches was used in combination with both the `even' and the `zero-or-two' encoding
 of the Brent equations. The resulting instances were presented to the SAT solver yalsat by \cite{yalsat}. When it didn't
 find a solution for an instance within a few minutes, the instance was discarded and a new instance with another
 random choice was tried. A detailed analysis of the effect of our optimizations on the performance of the solver
 is provided in a separate paper~\citep{sls}.

 \section{Recognizing Equivalences}\label{sec:sieving}

 From any given solution of the Brent equations we can generate many equivalent
 solutions.  For example, exchanging $\alpha$ with $\beta$ and flipping all
 indices maps a solution to another solution. This operation corresponds to the
 fact that $(AB)^\top=B^\top A^\top$.  It is also clear from the equations
 that replacing $\alpha$ by~$\beta$, $\beta$ by~$\gamma$, and $\gamma$
 by~$\alpha$ maps a solution to another solution, although this operation is
 less obvious in terms of matrix multiplication.  Finally, for any fixed
 invertible matrix~$U$, we can exploit the fact $AB=AUU^{-1}B$ to map solutions
 to other solutions.

 The operations just described form a group of symmetries of matrix
 multiplication which was introduced by \cite{de1978varieties}, who used them
 for showing that Strassen's scheme for $2\times 2$ matrices is essentially
 unique: it is unique modulo the action of this symmetry group. 
 To describe the group more formally, it is convenient to express matrix
 multiplication schemes as tensors,
 \[
 \sum_{\ell=1}^{23}
 \begin{pmatrix}
   \alpha_{1,1}\elldown & \alpha_{1,2}\elldown & \alpha_{1,3}\elldown \\
   \alpha_{2,1}\elldown & \alpha_{2,2}\elldown & \alpha_{2,3}\elldown \\
   \alpha_{3,1}\elldown & \alpha_{3,2}\elldown & \alpha_{3,3}\elldown
 \end{pmatrix}\otimes
 \begin{pmatrix}
   \beta_{1,1}\elldown & \beta_{1,2}\elldown & \beta_{1,3}\elldown \\
   \beta_{2,1}\elldown & \beta_{2,2}\elldown & \beta_{2,3}\elldown \\
   \beta_{3,1}\elldown & \beta_{3,2}\elldown & \beta_{3,3}\elldown
 \end{pmatrix}\otimes
 \begin{pmatrix}
   \gamma_{1,1}\elldown & \gamma_{1,2}\elldown & \gamma_{1,3}\elldown \\
   \gamma_{2,1}\elldown & \gamma_{2,2}\elldown & \gamma_{2,3}\elldown \\
   \gamma_{3,1}\elldown & \gamma_{3,2}\elldown & \gamma_{3,3}\elldown
 \end{pmatrix}.
 \]
 A scheme is correct if and only if it is equal, as element of $(K^{3\times
   3})^{\otimes3}$, to $\sum_{i,j,k=1}^3 E_{i,k}\otimes E_{k,j}\otimes E_{j,i}$,
 where $E_{u,v}\in K^{3\times3}$ refers to the matrix which has a $1$ at
 position $(u,v)$ and zeros everywhere else.

 A permutation $\pi\in S_3$ acts on
 a tensor $A\otimes B\otimes C$ by permuting the three factors, and transposing
 each of them if $\sgn(\pi)=-1$. For example, $(1\ 2)\cdot(A\otimes B\otimes
 C)=B^\top\otimes A^\top\otimes C^\top$ and $(1\ 2\ 3)\cdot(A\otimes B\otimes
 C)=B\otimes C\otimes A$. A triple $(U,V,W)\in\gl(K,3)^3$ of invertible
 matrices acts via
 \[
 (U,V,W)\cdot(A\otimes B\otimes C)=UAV^{-1}\otimes VBW^{-1}\otimes WCU^{-1}.
 \]
 A tuple $(U,V,W,\pi)\in\gl(K,3)^3\times S_3$ acts on a tensor $A\otimes
 B\otimes C$ by first letting the permutation act as described above, and then
 applying the matrices as described above.  The set $G=\gl(K,3)^3\times S_3$ is
 turned into a group by defining the multiplication in such a way that the
 operation described above becomes a group action.  The action of the group $G$
 defined on tensors $A\otimes B\otimes C$ is extended to the whole space
 $(K^{3\times 3})^{\otimes 3}$ by linearity. In other words, elements of $G$ act
 on sums of tensors by acting independently on all summands.

 Two matrix multiplication schemes are called equivalent if they belong to the
 same orbit under the action of~$G$. Whenever a new matrix multiplication scheme
 is discovered, the question is whether it is equivalent to a known scheme, for
 if it is, it should not be considered as new. A common test for checking that
 two schemes are not equivalent proceeds by computing certain invariants of the
 group action. For example, since permutation and multiplication by invertible
 matrices do not change the rank of a matrix, we can count how many matrices of
 rank 1,~2, and~3 appear in the scheme. If the counts differ for two schemes,
 then these schemes cannot be equivalent. For example, \cite{DBLP:journals/corr/abs-1108-2830}
 and \cite{oh2013inequivalence} proved in this way that their schemes were indeed
 new. Writing a scheme in the form $\sum_{\ell=1}^{23}(A_\ell\otimes
 B_\ell\otimes C_\ell)$, we can encode this invariant as the polynomial
 $\sum_{\ell=1}^{23} (x^{\rank(A_\ell)}+x^{\rank(B_\ell)}+x^{\rank(C_\ell)})$.
 Similarly, also the polynomials
 \[
   \sum_{\ell=1}^{23} x^{\rank(A_\ell)+\rank(B_\ell)+\rank(C_\ell)}
   \quad\text{and}\quad
   x^{\sum_{\ell=1}^{23}\rank(A_\ell)} +
   x^{\sum_{\ell=1}^{23}\rank(B_\ell)} +
   x^{\sum_{\ell=1}^{23}\rank(C_\ell)} 
 \]
 are invariants, because changing the order of summation does not affect the
 relative order of the factors in the tensor, and applying a permutation changes
 the relative order of the factors in every summand in the same way.

 When we have two schemes for which all three invariants match, they may
 nevertheless be inequivalent. For checking whether a solution found
 by the SAT solver is really new, comparing invariants is useful as a
 first step, but it is not sufficient. In fact, many solutions found
 by the SAT solver were inequivalent although all three invariants stated
 above agreed. Fortunately, it is not too hard to decide the equivalence of two given
 schemes by constructing, whenever possible, a group element that maps one to
 the other. We can proceed as follows. 

 Suppose we are given two
 multiplication schemes $S,S'$ and we want to decide whether there exists a
 tuple $(U,V,W,\pi)\in\gl(K,3)^3\times S_3$ such that $(U,V,W,\pi)\cdot S=S'$.
 As far as the permutation is concerned, there are only six candidates,
 so we can simply try each of them. Writing $S=\sum_{\ell=1}^{23}(A_\ell\otimes
 B_\ell\otimes C_\ell)$ and $S'=\sum_{\ell=1}^{23}(A_\ell'\otimes B_\ell'\otimes
 C_\ell')$, it remains to find $U,V,W$ that map all the summands of $S$ to the
 summands of~$S'$, albeit possibly in a different order. We search for a suitable
 order by the following recursive algorithm, which is initially called with
 $Q$ being full space $K^{3\times 3}\times K^{3\times 3}\times K^{3\times
   3}$.

 \medskip
 \par\noindent
 \emph{Input:} $S,S'$ as above, a basis of a subspace $Q$ of $K^{3\times 3}\times K^{3\times 3}\times K^{3\times 3}$\\
 \emph{Output:} A triple $(U,V,W)\in\gl(K,3)^3\cap Q$ with $(U,V,W)\cdot S=S'$, or $\bot$ if no such triple exists.

 \step10 if $S$ and $S'$ are empty, then:
 \step21 return any element $(U,V,W)$ of $Q$ with $\det(U)\det(V)\det(W)\neq0$, or $\bot$ if no such element exists.
 \step30 for all summands $A_\ell'\otimes B_\ell'\otimes C_\ell'$ of $S'$, do:
 \step41 if $\rank(A_1)=\rank(A_\ell')$ and $\rank(B_1)=\rank(B_\ell')$ and $\rank(C_1)=\rank(C_\ell')$, then:
 \step52 compute a basis of the space $P$ of all $(U,V,W)$ such that $UA_1=A_\ell'V$, $VB_1=B_\ell'W$, $WC_1=C_\ell'U$ by
    making an ansatz, comparing coefficients, and solving a homogeneous linear system.
 \step62 compute a basis of $R=P\cap Q$.
 \step72 if $R$ contains at least one triple $(U,V,W)$ with $\det(U)\det(V)\det(W)\neq0$, then:
 \step83 call the algorithm recursively with the first summand of $S$ and the $\ell$th summand of $S'$ removed, and with $R$ in place of~$Q$.
 \step93 if the recursive call yields a triple $(U,V,W)$, return it.
 \step{10}0 return $\bot$.

 \medskip
 The algorithm terminates because each recursive call is applied to a sum with
 strictly fewer summands.
 The correctness of the algorithm is clear because it essentially performs an
 exhaustive search through all options. 
 In order to perform the check in
 step~7, we can consider a generic linear combination of the basis elements
 of~$R$, with variables as coefficients. Then $\det(U)\det(V)\det(W)$ is a
 polynomial in these variables, and the question is whether this polynomial
 vanishes identically on~$K$. Since we are interested in the case $K=\set Z_2$,
 we can answer this by an exhaustive search.

 The recursive structure of the algorithm with up to 23 recursive calls at every
 level may seem prohibitively expensive.  However, the two filters in lines~4
 and~7 turn out to cut down the number of recursive calls considerably. A
 straightforward implementation in Mathematica needs no more than about one
 second of computation time to decide whether or not two given schemes are
 equivalent. Of course, we first compare the invariants, which is almost for
 free and suffices to settle many cases.

 For each scheme found by the SAT solver we have checked whether it is
 equivalent (for $K=\set Z_2$) to one of the schemes found earlier, or to one
 of the four known schemes found by Laderman, Smirnov, Oh et al., and
 Courtois et al., respectively. From the roughly $270\,000$ solutions
 found by the SAT solver that were distinct modulo the order of the summands,
 we isolated about $13\,000$ schemes that were distinct modulo equivalence.
 In the appendix, we list the number of schemes we found separated by
 invariant. 

 \section{Simplifying Solutions}\label{sec:switching}

 We can use the symmetries introduced in the previous section not only to
 recognize that a seemingly new scheme is not really new.  We can also use
 them for simplifying schemes. A scheme can for example be regarded
 as simpler than another scheme if the number of terms
 $\alpha_{i_1,i_2}\elldown\beta_{j_1,j_2}\elldown\gamma_{k_1,k_2}\elldown$ in
 it which evaluate to $1$ is smaller. Calling this number the \emph{weight} of
 a scheme, we prefer schemes with smaller weight.

 Ideally, we would like to replace every scheme $S$ by an equivalent scheme
 with smallest possible weight. In principle, we could find such a minimal
 equivalent element by applying all elements of $G$ to $S$ and taking the
 smallest result. Unfortunately, even for $K=\set Z_2$, the group $G$ has
 $168^3\cdot6=28\,449\,792$ elements, so trying them all might be feasible if we had
 to do it for a few schemes, but not for thousands of them.
 If we do not insist in the smallest possible weight, we can take a pragmatic
 approach and just spend for every scheme $S$ a prescribed amount of computation
 time (say half an hour) applying random elements of $G$ to~$S$:

 \medskip
 \par\noindent
 \emph{Input:} a multiplication scheme $S$\\
 \emph{Output:} an equivalent multiplication scheme whose weight is less than or equal to the weight of~$S$.

 \step10 while the time limit is not exhausted, do
 \step21 pick a group element $g$ at random
 \step31 if $\mathrm{weight}(g(S))<\mathrm{weight}(S)$, then set $S = g(S)$
 \step40 return $S$

 \medskip
 With this algorithm, we were able to replace about 20\% of the new schemes found by the SAT solver
 by equivalent schemes with smaller weight. It is not too surprising that no
 improvement was found for the majority of cases, because the way we specified the
 problem to the SAT solver already induces a bias towards solutions with a small weight.

 The figure below shows the distribution of our $13\,000$ schemes according to
 weight, after simplification. It is clear that the weight is always odd, hence
 the small gaps between the bars. It is less clear why we seem to have an
 overlay of three normal distributions, but we believe that this is rather an
 artifact of the way we generated the solutions than a structural feature of the
 entire solution set.

 
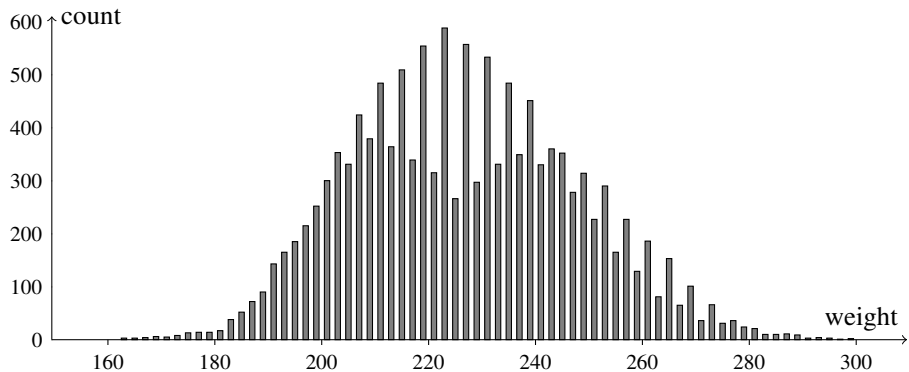
\begin{center}
 \begin{tikzpicture}[x=2pt,y=2pt,yscale=.1,xscale=1]
\foreach \x/\y in {257/227,259/129,261/186,263/81,265/153,267/65,269/101,271/36,273/66,167/4,275/31,277/36,279/24,281/21,283/10,285/10,287/11,289/9,163/3,293/4,295/3,297/1,299/2,173/8,175/13,177/14,179/14,181/17,183/38,185/52,187/72,189/90,169/6,191/143,193/165,195/185,197/215,199/252,201/300,203/353,205/331,291/3,207/424,209/379,211/484,213/364,215/509,217/339,219/554,221/315,223/588,225/266,227/557,229/297,165/3,231/533,233/331,171/5,235/484,237/349,239/451,241/330,243/360,245/352,247/278,249/314,251/227,253/290,255/165}
\draw[fill=gray] (\x,0) rectangle (\x+1,\y);
\draw[->] (150,0)--(310,0) node[above left] {weight};
\draw[->] (150,0)--++(0,610) node[right] {count};
\foreach \x in {160,180,200,220,240,260,280,300} \draw (\x.5,0)--++(0,-10) node[below] {\footnotesize\x};
\foreach \y in {0,100,200,300,400,500,600} \draw (150,\y)--++(0,-1) node[left] {\footnotesize\y};
 \end{tikzpicture}
 \end{center}
 
 \section{Generalizing the Coefficient Ring}\label{sec:signing}

 At this point, we have a considerable number of new matrix multiplication
 schemes for the coefficient field $K=\set Z_2$. The next step is to lift them
 to schemes that work in any coefficient ring. 
 The SAT solver presents us with a solution for $\set Z_2$ in which all
 coefficients are $0$ or~$1$, and in order to lift such a solution, we make the
 hypothesis that this solution originated from a solution for an arbitrary coefficient
 ring in which all coefficients are $+1$, $-1$, or~$0$. The distinction between
 $+1$ and $-1$ gets lost in~$\set Z_2$, and the task consists in recovering it.
 There is a priori no reason why such a lifting should exist, and indeed, we have
 seen a small number of instances where it fails. One such example is given in the
 appendix. Interestingly however, these examples seem to be very rare. In almost all cases,
 a lifting turned out to exist. 

 In order to explain the lifting process, we return to the Brent equations discussed in
 Section~\ref{sec:brent}. We set variables corresponding to coefficients
 that are zero in the SAT solution to zero, which simplifies the system
 considerably. According to the axioms of
 tensor products, we have $(\lambda A)\otimes B\otimes C=A\otimes(\lambda B)
 \otimes C=A\otimes B\otimes(\lambda C)$ for any $A,B,C$ and every
 constant~$\lambda$. We may therefore select in every summand $A\otimes
 B\otimes C$ one variable appearing in $A$ and one variable appearing in $B$ and
 set them to~$+1$. This reduces the number of variables further. However, the
 resulting system is still to hard to be solved directly.

 Before calling a general
 purpose Gr\"obner bases engine, we apply some simplifications to the system,
 which take into account that we are only interested in solutions whose
 coordinates are $-1$ or~$+1$. In particular, we can replace any exponent $k$
 appearing in any of the polynomials by~$k\bmod 2$, we can cancel factors that
 clearly do not vanish on the points of interest, and we can replace polynomials
 of the from $xy\pm1$ by $x\pm y$.
 These simplifications may bring up some linear polynomials. By triangularizing the linear system
 corresponding to these polynomials, we can eliminate some of the variables. We can then
 simplify again, and possibly obtain new linear equations. The process is repeated until
 no further linear equations appear. We then add for each variable $x$ the polynomial $x^2-1$
 and compute a Gr\"obner basis with respect to a degree order. If this leads to new
 linear polynomials, we return to iterating triangularization, elimination,
 and simplification until no further linear equations show up, and then compute again a
 degree Gr\"obner basis. The whole process is repeated until we obtain a Gr\"obner basis
 that does not contain any new linear equations. If there are more than 15 variables left,
 we next compute a minimal associated prime ideal of an elimination ideal involving only five
 variables, and check whether adding it to the original system and computing a Gr\"obner basis
 leads to new linear equations. If it does, we start over with the whole procedure.
 Otherwise, we compute the minimal associated prime ideal of the whole system and return
 the solution corresponding to one of the prime factors. The process is summarized in the following
 listing.

 \medskip
 \par\noindent
 \emph{Input:} A finite subset $B$ of $\set Q[x_1,\dots,x_n]$\\
 \emph{Output:} A common root $\xi\in\{-1,1\}^n$ of all the elements of~$B$, or $\bot$ if no such common root exists.

 \step10 Replace every exponent $k$ appearing in an element of $B$ by $k\bmod2$
 \step20 For every $p\in B$ and every $i$ with $x_i\mid p$, replace $p$ by $p/x_i$
 \step30 Replace every element of the form $xy-1$ or $-xy-1$ by $x-y$ or $x+y$, respectively.
 \step40 if $B$ now contains linear polynomials, then:
 \step51 Use them to eliminate some variables, say $y_1,\dots,y_k$
 \step61 Call the procedure recursively on the resulting set of polynomials
 \step71 if there is a solution, extend it to the eliminated variables $y_1,\dots,y_k$ and return the result
 \step81 if there is no solution, return $\bot$.
 \step{9}0 Compute a Gr\"obner basis $G$ of $B\cup\{x_i^2-1:i=1,\dots,n\}$ with respect to a degree order
 \step{10}0 if $G=\{1\}$, return $\bot$
 \step{11}0 if $G$ contains linear polynomials, then call this procedure recursively and return the result
 \step{12}0 if $n>15$, then:
 \step{13}1 Compute a basis $P$ of one of the minimal associated prime ideals of $\<G>\cap\set Q[x_1,\dots,x_5]$.
 \step{14}1 Compute a Gr\"obner basis $G'$ of $G\cup P$ with respect to a degree order
 \step{15}1 if $G'$ contains linear polynomials, then call this procedure recursively and return the result
 \step{16}0 Compute a basis $P$ of one of the minimal associated prime ideals of $\<G>\subseteq\set Q[x_1,\dots,x_n]$.
 \step{17}0 Return the common solution $\xi$ of~$P$.

 \medskip
 An implementation of this procedure in Mathematica is available on the website of this
 article~\citep{mmr}. In this implementation, we use Singular \citep{greuel02} for
 doing the Gr\"obner basis calculations and for the computation of minimal associated
 prime ideals. Despite the large number of
 variables, Singular handles the required computations with impressive
 speed, so that the whole signing process takes only about 20 seconds per solution on
 the average. Only a small number of cases, which happen to have a few more
 variables than the others, need much longer, up to a few hours.

 \section{Introducing Parameters}\label{sec:families}

 The idea of instantiating some of the variables based on a known scheme and
 then solving for the remaining variables approach not only applies to SAT
 solving. It also has an algebraic counterpart. Solving the Brent equations with
 algebraic methods is infeasible because the equations are nonlinear, but
 observe that we only have to solve a linear system if we start from a known
 scheme and only replace all $\gamma_{k_1,k_2}\elldown$ by fresh
 variables. Solving linear systems is of course much easier than
 solving nonlinear ones.

 More generally, we can select for each $\ell\in\{1,\dots,23\}$ separately
 whether we want to replace all $\alpha_{i_1,i_2}\elldown$'s or all
 $\beta_{j_1,j_2}\elldown$'s or all $\gamma_{k_1,k_2}\elldown$'s by fresh
 variables, and we still just get a linear system for these variables. Once we
 make a selection, solving the resulting linear system yields an affine vector
 space. One might expect this affine space will typically consist of a single
 point only, but this is usually not the case.

 A solution space with positive dimension can be translated into a
 multiplication scheme involving one or more free parameters. Starting from the
 resulting parameterized scheme, we can play the same game with another
 selection of variables, which may allow us to introduce further parameters. If
 we repeat the procedure several times with random selections of which variables
 are known, we obtain huge schemes involving 40 or more parameters.
 These parameters are however algebraically dependent, or at least it is too
 costly check whether they are dependent or not. We got better results by
 proceeding more systematically, as summarized int in the following listing.

 \medskip
 \par\noindent
 \emph{Input:} A matrix multiplication scheme $S=\sum_{\ell=1}^{23}(A_\ell\otimes B_\ell\otimes C_\ell)$.
 Write $A_\ell=((\alpha_{i,j}\elldown))$, $B_\ell=((\beta_{i,j}\elldown))$, $C_\ell=((\gamma_{i,j}\elldown))$.\\
 \emph{Output:} A family of matrix multiplication schemes with parameters $x_1,x_2,\dots$

 \step10 for $\ell=1,\dots,23$, do:
 \step21 for every choice $u,v\in\{\alpha,\beta,\gamma\}$ with $u\neq v$, do:
 \step32 replace all entries $u_{i,j}\elldown$ for $i,j=1,\dots,3$ in $S$ by fresh variables
 \step42 replace all entries $v_{i,j}^{(m)}$ for $i,j=1,\dots,3$ and $m\neq\ell$ in $S$ by fresh variables
 \step52 equate the resulting scheme $S$ to $\sum_{i,j,k} E_{i,j}\otimes E_{j,k}\otimes E_{k,i}$ and compare coefficients
 \step62 solve the resulting inhomogeneous linear system for the fresh variables introduced in steps 3 and~4
 \step72 substitute the generic solution, using new parameters $x_i,x_{i+1},\dots$, into~$S$
 \step80 return $S$

 \medskip
 With this algorithm and some slightly modified variants (e.g., letting the outer loop run backwards or transposing
 the inner and the outer loop), we were able to obtain schemes with altogether up to 17 parameters.
 Although all new parameters introduced in a certain iteration can only appear
 linearly in the scheme, old parameters that were considered as belonging to the
 ground ring during the linear solving can later appear rationally. However, by manually
 applying suitable changes of variables, we managed to remove all denominators from
 all the families we inspected. Not even integer denominators are needed. 
 We can also check using Gr\"obner bases
 whether the parameters are independent, and for several families with 17 parameters
 they turn out to be. In the language of algebraic geometry, this means that the solution
 set of the Brent equations has at least dimension~17 as an algebraic variety.

 One of our families is shown in the appendix, and some further ones are provided
 electronically on our website. These families should be contrasted with the family found by
 Johnson and McLoughlin in the the 1980s~\citep{DBLP:journals/siamcomp/JohnsonM86}. In particular, while they lament
 that their family contains fractional coefficients such as $\frac12$ and $\frac13$
 and therefore does not apply in every coefficient ring, our families only involve
 integer coefficients and therefore have no such restriction. Moreover, their family
 has only three parameters, and with the method described above, only $6$ additional
 parameters can be introduced into it. The number of parameters we managed to introduce
 into the known solutions by Laderman, Courtois et al., Oh et al., and Smirnov
 are $0$,~$6$, $10$, and~$14$, respectively.
 
 \section{Concluding Remarks}\label{sec:skimming}

 Although we have found many new multiplication schemes with 23 multiplications,
 we did not encounter a single scheme with 22 multiplications. We have checked
 all schemes whether some of their summands can be merged together using tensor product
 arithmetic. For doing so, it would suffice if a certain scheme contains some summands
 which share the same $A$'s, say, and where the
 corresponding $B$'s, say, of these rows are linearly independent. We could then
 express one of these $B$'s in terms of the others and eliminate the summand in
 which it appears. For example, if $B_3=\beta_1B_1+\beta_2B_2$, then we have $A\otimes B_1\otimes C_1
 + A\otimes B_2\otimes C_2+A\otimes B_3\otimes C_3=A\otimes
 B_1\otimes(C_1+\beta_1C_3)+A\otimes B_2\otimes(C_2+\beta_2C_3)$. Since none of
 our schemes admits a simplification of this kind, it remains open whether a
 scheme with 22 multiplications exists.

 Another open question is: how many further schemes with 23 multiplications and coefficients in $\{-1,0,1\}$
 are there? We have no evidence that we have found them all. In fact, we rather believe that there
 are many further ones, possibly including schemes that are very different from
 ours. There may also be parametrized families with more than 17 parameters, and it would
 be interesting to know the maximal possible number of parameters, i.e., the actual dimension
 of the solution set of the Brent equations.  

\bibliographystyle{elsarticle-harv}
 \bibliography{all}

\begin{thebibliography}{36}
\expandafter\ifx\csname natexlab\endcsname\relax\def\natexlab#1{#1}\fi
\expandafter\ifx\csname url\endcsname\relax
  \def\url#1{\texttt{#1}}\fi
\expandafter\ifx\csname urlprefix\endcsname\relax\def\urlprefix{URL }\fi

\bibitem[{Biere(2018)}]{yalsat}
Biere, A., 2018. {CaDiCaL, Lingeling, Plingeling, Treengeling and YalSAT
  Entering the SAT Competition 2018}. In: Proc.~of {SAT Competition} 2018 --
  Solver and Benchmark Descriptions. Vol. B-2018-1 of Department of Computer
  Science Series of Publications B. University of Helsinki, pp. 13--14.

\bibitem[{Biere et~al.(2009)Biere, Heule, van Maaren, and
  Walsh}]{DBLP:series/faia/2009-185}
Biere, A., Heule, M., van Maaren, H., Walsh, T. (Eds.), 2009. Handbook of
  Satisfiability. Vol. 185 of Frontiers in Artificial Intelligence and
  Applications. {IOS} Press.

\bibitem[{Bl{\"a}ser(2003)}]{blaser2003complexity}
Bl{\"a}ser, M., 2003. On the complexity of the multiplication of matrices of
  small formats. Journal of Complexity 19~(1), 43--60.

\bibitem[{Bl{\"a}ser(2013)}]{blaeser-survey2013}
Bl{\"a}ser, M., 2013. Fast Matrix Multiplication. No.~5 in Graduate Surveys.
  Theory of Computing Library.
\newline\urlprefix\url{http://www.theoryofcomputing.org/library.html}

\bibitem[{Brent(1970)}]{brent1970algorithms}
Brent, R.~P., 1970. Algorithms for matrix multiplication. Tech. rep.,
  Department of Computer Science, Stanford.

\bibitem[{Buchberger(1965)}]{buchberger65}
Buchberger, B., 1965. {Ein Algorithmus zum Auffinden der Basiselemente des
  Restklassenrings nach einem nulldimensionalen Polynomideal}. Ph.D. thesis,
  Universit{\"a}t Innsbruck.

\bibitem[{Buchberger and Kauers(2010)}]{buchberger10}
Buchberger, B., Kauers, M., 2010. Gr{\"o}bner basis. Scholarpedia 5~(10), 7763,
  \url{http://www.scholarpedia.org/article/Groebner_basis}.

\bibitem[{B{\"u}rgisser et~al.(2013)B{\"u}rgisser, Clausen, and
  Shokrollahi}]{buergisser-book}
B{\"u}rgisser, P., Clausen, M., Shokrollahi, M.~A., 2013. Algebraic complexity
  theory. Vol. 315. Springer Science \& Business Media.

\bibitem[{Courtois et~al.(2011)Courtois, Bard, and
  Hulme}]{DBLP:journals/corr/abs-1108-2830}
Courtois, N., Bard, G.~V., Hulme, D., 2011. A new general-purpose method to
  multiply $3\times3$ matrices using only 23 multiplications. CoRR
  abs/1108.2830.
\newline\urlprefix\url{http://arxiv.org/abs/1108.2830}

\bibitem[{Cox et~al.(1992)Cox, Little, and OShea}]{cox2013ideals}
Cox, D., Little, J., OShea, D., 1992. Ideals, Varieties, and Algorithms.
  Undergraduate Texts in Mathematics. Springer.

\bibitem[{de~Groote(1978)}]{de1978varieties}
de~Groote, H.~F., 1978. On varieties of optimal algorithms for the computation
  of bilinear mappings i. the isotropy group of a bilinear mapping. Theoretical
  Computer Science 7~(1), 1--24.

\bibitem[{Drevet et~al.(2011)Drevet, Islam, and
  Schost}]{DBLP:journals/tcs/DrevetIS11}
Drevet, C., Islam, M.~N., Schost, {\'{E}}., 2011. Optimization techniques for
  small matrix multiplication. Theor. Comput. Sci. 412~(22), 2219--2236.

\bibitem[{Dumas et~al.(2008)Dumas, Giorgi, and Pernet}]{DGP:2008}
Dumas, J.-G., Giorgi, P., Pernet, C., 2008. Dense linear algebra over word-size
  prime fields: the fflas and ffpack packages. ACM Trans. on Mathematical
  Software (TOMS) 35~(3), 1--42.

\bibitem[{Gomes and Sellmann(2004)}]{streamlining}
Gomes, C., Sellmann, M., 2004. Streamlined constraint reasoning. In: Principles
  and Practice of Constraint Programming (CP 2004). Springer Berlin Heidelberg,
  Berlin, Heidelberg, pp. 274--289.

\bibitem[{Greuel and Pfister(2002)}]{greuel02}
Greuel, G.-M., Pfister, G., 2002. A {S}ingular Introduction to Commutative
  Algebra. Springer.

\bibitem[{Heule et~al.(2019{\natexlab{a}})Heule, Kauers, and Seidl}]{sls}
Heule, M.~J., Kauers, M., Seidl, M., 2019{\natexlab{a}}. {Local Search for Fast
  Matrix Multiplication}. In: Proceedings of SAT'19. To appear; also ArXiv
  1903.11391.

\bibitem[{Heule et~al.(2019{\natexlab{b}})Heule, Kauers, and Seidl}]{mmr}
Heule, M.~J., Kauers, M., Seidl, M., 2019{\natexlab{b}}. Matrix multiplication
  repository.
  \url{http://www.algebra.uni-linz.ac.at/research/matrix-multiplication/}.

\bibitem[{Heule(2018)}]{DBLP:conf/aaai/Heule18}
Heule, M. J.~H., 2018. Schur number five. In: McIlraith, S.~A., Weinberger,
  K.~Q. (Eds.), Proceedings of the Thirty-Second {AAAI} Conference on
  Artificial Intelligence, (AAAI-18), the 30th innovative Applications of
  Artificial Intelligence (IAAI-18), and the 8th {AAAI} Symposium on
  Educational Advances in Artificial Intelligence (EAAI-18). {AAAI} Press, pp.
  6598--6606.
\newline\urlprefix\url{https://www.aaai.org/ocs/index.php/AAAI/AAAI18/paper/view/16952}

\bibitem[{Heule et~al.(2016)Heule, Kullmann, and
  Marek}]{DBLP:conf/sat/HeuleKM16}
Heule, M. J.~H., Kullmann, O., Marek, V.~W., 2016. Solving and verifying the
  boolean {P}ythagorean triples problem via cube-and-conquer. In: Creignou, N.,
  Berre, D.~L. (Eds.), Proceedings of the 19th International Conference on
  Theory and Applications of Satisfiability Testing ({SAT} 2016). Vol. 9710 of
  Lecture Notes in Computer Science. Springer, pp. 228--245.
\newline\urlprefix\url{https://doi.org/10.1007/978-3-319-40970-2_15}

\bibitem[{Huang et~al.(2016)Huang, Smith, Henry, and van~de
  Geijn}]{Huang:2016:SAR:3014904.3014983}
Huang, J., Smith, T.~M., Henry, G.~M., van~de Geijn, R.~A., 2016. Strassen's
  algorithm reloaded. In: Proceedings of the International Conference for High
  Performance Computing, Networking, Storage and Analysis. SC'16. IEEE Press,
  Piscataway, NJ, USA, pp. 59:1--59:12.
\newline\urlprefix\url{http://dl.acm.org/citation.cfm?id=3014904.3014983}

\bibitem[{Johnson and McLoughlin(1986)}]{DBLP:journals/siamcomp/JohnsonM86}
Johnson, R.~W., McLoughlin, A.~M., 1986. Noncommutative bilinear algorithms for
  $3\times3$ matrix multiplication. {SIAM} J. Comput. 15~(2), 595--603.
\newline\urlprefix\url{https://doi.org/10.1137/0215043}

\bibitem[{Laderman(1976)}]{laderman1976noncommutative}
Laderman, J.~D., 1976. A noncommutative algorithm for multiplying $3\times3$
  matrices using 23 multiplications. Bulletin of the American Mathematical
  Society 82~(1), 126--128.

\bibitem[{Landsberg(2017)}]{landsberg2017geometry}
Landsberg, J.~M., 2017. Geometry and complexity theory. Vol. 169. Cambridge
  University Press.

\bibitem[{Le~Gall(2014)}]{LeGall:2014:PTF:2608628.2608664}
Le~Gall, F., 2014. Powers of tensors and fast matrix multiplication. In:
  Proceedings of the 39th International Symposium on Symbolic and Algebraic
  Computation. ISSAC'14. ACM, pp. 296--303.
\newline\urlprefix\url{http://doi.acm.org/10.1145/2608628.2608664}

\bibitem[{Makarov(1987)}]{MAKAROV1987205}
Makarov, O., 1987. A non-commutative algorithm for multiplying $5\times5$
  matrices using one hundred multiplications. USSR Computational Mathematics
  and Mathematical Physics 27~(1), 205 -- 207.
\newline\urlprefix\url{http://www.sciencedirect.com/science/article/pii/0041555387901455}

\bibitem[{Oh et~al.(2013)Oh, Kim, and Moon}]{oh2013inequivalence}
Oh, J., Kim, J., Moon, B.-R., 2013. On the inequivalence of bilinear algorithms
  for $3\times3$ matrix multiplication. Information Processing Letters
  113~(17), 640--645.

\bibitem[{Pan(2018)}]{pan-survey2018}
Pan, V.~Y., 2018. Fast feasible and unfeasible matrix multiplication. CoRR
  abs/1804.04102.
\newline\urlprefix\url{http://arxiv.org/abs/1804.04102}

\bibitem[{Sedoglavic(2017{\natexlab{a}})}]{DBLP:journals/corr/Sedoglavic17}
Sedoglavic, A., 2017{\natexlab{a}}. Laderman matrix multiplication algorithm
  can be constructed using strassen algorithm and related tensor's isotropies.
  CoRR abs/1703.08298.
\newline\urlprefix\url{http://arxiv.org/abs/1703.08298}

\bibitem[{Sedoglavic(2017{\natexlab{b}})}]{DBLP:journals/corr/Sedoglavic17aa}
Sedoglavic, A., 2017{\natexlab{b}}. A non-commutative algorithm for multiplying
  $5\times5$ matrices using 99 multiplications. CoRR abs/1707.06860.
\newline\urlprefix\url{http://arxiv.org/abs/1707.06860}

\bibitem[{Sedoglavic(2017{\natexlab{c}})}]{DBLP:journals/corr/abs-1712-07935}
Sedoglavic, A., 2017{\natexlab{c}}. A non-commutative algorithm for multiplying
  $7\times7$ matrices using 250 multiplications. CoRR abs/1712.07935.
\newline\urlprefix\url{http://arxiv.org/abs/1712.07935}

\bibitem[{Sedoglavic(2019)}]{fastmm}
Sedoglavic, A., 2019. {Yet another catalogue of fast matrix multiplication
  algorithms}. \url{https://fmm.univ-lille.fr/}, accessed: 2019-03-17.

\bibitem[{Smirnov(2013)}]{smirnov2013bilinear}
Smirnov, A.~V., 2013. The bilinear complexity and practical algorithms for
  matrix multiplication. Computational Mathematics and Mathematical Physics
  53~(12), 1781--1795.

\bibitem[{Smirnov(2017)}]{smirnovseveral}
Smirnov, A.~V., 2017. Several bilinear algorithms for matrix multiplication.
  Tech. rep., Technical report.

\bibitem[{Strassen(1969)}]{strassen1969gaussian}
Strassen, V., 1969. Gaussian elimination is not optimal. Numerische Mathematik
  13~(4), 354--356.

\bibitem[{Williams(2012)}]{Williams:2012:MMF:2213977.2214056}
Williams, V.~V., 2012. Multiplying matrices faster than
  {C}oppersmith-{W}inograd. In: Proceedings of the 44th Annual ACM Symposium on
  Theory of Computing. STOC'12. ACM, New York, NY, USA, pp. 887--898.
\newline\urlprefix\url{http://doi.acm.org/10.1145/2213977.2214056}

\bibitem[{Winograd(1971)}]{winograd1971multiplication}
Winograd, S., 1971. On multiplication of $2\times2$ matrices. Linear algebra
  and its applications 4~(4), 381--388.

\end{thebibliography}

 \section*{Appendix}

\noindent 
$\blacksquare$\quad List of all invariants appearing in our set of non-equivalent schemes. These tables are based on the schemes
for $\set Z_2$ and include the few schemes that could not be lifted to~$\set Z$. The numbers on the right indicate how many schemes
with the respective invariant we have found.

\[
 \sum_{\ell=1}^{23} (x^{\rank(A_\ell)}+x^{\rank(B_\ell)}+x^{\rank(C_\ell)})
\]
{\footnotesize

\begin{tabular}{r|l}
$ 14 x^{2} + 55 x $ & 1 \\
$ 2 x^{3} + 25 x^{2} + 42 x $ & 1 \\
$ 6 x^{3} + 12 x^{2} + 51 x $ & 1 \\
$ 4 x^{3} + 18 x^{2} + 47 x $ & 2 \\
$ 2 x^{3} + 24 x^{2} + 43 x $ & 3 \\
$ 4 x^{3} + 21 x^{2} + 44 x $ & 3 \\
$ 2 x^{3} + 21 x^{2} + 46 x $ & 3 \\
$ 2 x^{3} + 23 x^{2} + 44 x $ & 3 \\
$ x^{3} + 27 x^{2} + 41 x $ & 4 \\
$ 2 x^{3} + 22 x^{2} + 45 x $ & 4 \\
$ 4 x^{3} + 19 x^{2} + 46 x $ & 5 \\
\end{tabular}\hfill\begin{tabular}{r|l}
$ 4 x^{3} + 20 x^{2} + 45 x $ & 5 \\
$ 25 x^{2} + 44 x $ & 6 \\
$ 2 x^{3} + 20 x^{2} + 47 x $ & 7 \\
$ 24 x^{2} + 45 x $ & 7 \\
$ x^{3} + 26 x^{2} + 42 x $ & 8 \\
$ x^{3} + 16 x^{2} + 52 x $ & 10 \\
$ 3 x^{3} + 16 x^{2} + 50 x $ & 16 \\
$ x^{3} + 25 x^{2} + 43 x $ & 17 \\
$ 23 x^{2} + 46 x $ & 26 \\
$ 3 x^{3} + 22 x^{2} + 44 x $ & 31 \\
$ x^{3} + 17 x^{2} + 51 x $ & 47 \\
\end{tabular}\hfill\begin{tabular}{r|l}
$ x^{3} + 24 x^{2} + 44 x $ & 47 \\
$ 3 x^{3} + 17 x^{2} + 49 x $ & 53 \\
$ 15 x^{2} + 54 x $ & 58 \\
$ 3 x^{3} + 21 x^{2} + 45 x $ & 68 \\
$ x^{3} + 23 x^{2} + 45 x $ & 77 \\
$ 3 x^{3} + 18 x^{2} + 48 x $ & 102 \\
$ 3 x^{3} + 20 x^{2} + 46 x $ & 103 \\
$ x^{3} + 22 x^{2} + 46 x $ & 104 \\
$ 16 x^{2} + 53 x $ & 110 \\
$ x^{3} + 18 x^{2} + 50 x $ & 111 \\
$ x^{3} + 21 x^{2} + 47 x $ & 138 \\
\end{tabular}\hfill\begin{tabular}{r|l}
$ x^{3} + 19 x^{2} + 49 x $ & 139 \\
$ 3 x^{3} + 19 x^{2} + 47 x $ & 144 \\
$ x^{3} + 20 x^{2} + 48 x $ & 147 \\
$ 22 x^{2} + 47 x $ & 375 \\
$ 17 x^{2} + 52 x $ & 914 \\
$ 21 x^{2} + 48 x $ & 1371 \\
$ 18 x^{2} + 51 x $ & 2461 \\
$ 20 x^{2} + 49 x $ & 2824 \\
$ 19 x^{2} + 50 x $ & 3476 \\
\vphantom{$ 19 x^{2} + 50 x $} & \vphantom{3476} \\
\vphantom{$ 19 x^{2} + 50 x $} & \vphantom{3476} \\
\end{tabular}

}
\[
\sum_{\ell=1}^{23} x^{\rank(A_\ell)+\rank(B_\ell)+\rank(C_\ell)}
\]
{\footnotesize   

  \begin{tabular}{r|l}
$ x^{7} + 3 x^{6} + 3 x^{5} + 5 x^{4} + 11 x^{3} $ & 1 \\
$ x^{7} + 3 x^{6} + 3 x^{5} + 4 x^{4} + 12 x^{3} $ & 1 \\
$ 5 x^{6} + 2 x^{5} + 7 x^{4} + 9 x^{3} $ & 1 \\
$ 4 x^{6} + 6 x^{5} + 13 x^{3} $ & 1 \\
$ 4 x^{6} + 2 x^{4} + 17 x^{3} $ & 1 \\
$ 4 x^{6} + x^{5} + 3 x^{4} + 15 x^{3} $ & 1 \\
$ 4 x^{6} + 5 x^{5} + 2 x^{4} + 12 x^{3} $ & 1 \\
$ 4 x^{6} + 6 x^{5} + 3 x^{4} + 10 x^{3} $ & 1 \\
$ 4 x^{6} + 3 x^{5} + 2 x^{4} + 14 x^{3} $ & 1 \\
$ 5 x^{6} + 2 x^{5} + 6 x^{4} + 10 x^{3} $ & 1 \\
$ 4 x^{6} + 5 x^{5} + 7 x^{4} + 7 x^{3} $ & 2 \\
$ 4 x^{6} + 6 x^{5} + 5 x^{4} + 8 x^{3} $ & 2 \\
$ 4 x^{6} + 3 x^{5} + 10 x^{4} + 6 x^{3} $ & 2 \\
$ 4 x^{6} + 3 x^{5} + 4 x^{4} + 12 x^{3} $ & 3 \\
$ 3 x^{6} + 4 x^{5} + 5 x^{4} + 11 x^{3} $ & 3 \\
$ 4 x^{6} + 5 x^{5} + 4 x^{4} + 10 x^{3} $ & 3 \\
$ 3 x^{6} + 5 x^{5} + 5 x^{4} + 10 x^{3} $ & 3 \\
$ 5 x^{6} + 2 x^{5} + 5 x^{4} + 11 x^{3} $ & 3 \\
$ 4 x^{6} + 3 x^{5} + 9 x^{4} + 7 x^{3} $ & 3 \\
$ 3 x^{6} + 4 x^{5} + 7 x^{4} + 9 x^{3} $ & 3 \\
    $ 4 x^{6} + 3 x^{5} + 11 x^{4} + 5 x^{3} $ & 4
\end{tabular}\hfill\begin{tabular}{r|l}
$ 4 x^{6} + x^{5} + 2 x^{4} + 16 x^{3} $ & 4 \\
$ 4 x^{6} + 5 x^{5} + 6 x^{4} + 8 x^{3} $ & 5 \\
$ 3 x^{6} + 5 x^{5} + 4 x^{4} + 11 x^{3} $ & 5 \\
$ 3 x^{6} + 4 x^{5} + 4 x^{4} + 12 x^{3} $ & 5 \\
$ 4 x^{6} + x^{5} + 9 x^{4} + 9 x^{3} $ & 5 \\
$ 4 x^{6} + 3 x^{5} + 7 x^{4} + 9 x^{3} $ & 5 \\
$ 4 x^{6} + 5 x^{5} + 5 x^{4} + 9 x^{3} $ & 5 \\
$ 4 x^{6} + 3 x^{5} + 8 x^{4} + 8 x^{3} $ & 5 \\
$ 4 x^{6} + 2 x^{5} + 12 x^{4} + 5 x^{3} $ & 6 \\
$ 3 x^{6} + 4 x^{5} + 8 x^{4} + 8 x^{3} $ & 6 \\
$ 4 x^{6} + 3 x^{5} + 5 x^{4} + 11 x^{3} $ & 8 \\
$ 4 x^{6} + 2 x^{5} + 2 x^{4} + 15 x^{3} $ & 8 \\
$ 4 x^{6} + 3 x^{5} + 6 x^{4} + 10 x^{3} $ & 9 \\
$ 4 x^{6} + 2 x^{5} + 11 x^{4} + 6 x^{3} $ & 14 \\
$ 4 x^{6} + 4 x^{5} + 2 x^{4} + 13 x^{3} $ & 16 \\
$ 4 x^{6} + 4 x^{5} + 8 x^{4} + 7 x^{3} $ & 34 \\
$ 4 x^{6} + 2 x^{5} + 10 x^{4} + 7 x^{3} $ & 42 \\
$ 4 x^{6} + 2 x^{5} + 3 x^{4} + 14 x^{3} $ & 43 \\
$ 4 x^{6} + 4 x^{5} + 3 x^{4} + 12 x^{3} $ & 53 \\
$ 4 x^{6} + 3 x^{4} + 16 x^{3} $ & 58 \\
  $ 4 x^{6} + 4 x^{5} + 7 x^{4} + 8 x^{3} $ & 70
\end{tabular}\hfill\begin{tabular}{r|l}  
$ 4 x^{6} + 10 x^{4} + 9 x^{3} $ & 71 \\
$ 4 x^{6} + 2 x^{5} + 9 x^{4} + 8 x^{3} $ & 72 \\
$ 4 x^{6} + 2 x^{5} + 8 x^{4} + 9 x^{3} $ & 95 \\
$ 4 x^{6} + 4 x^{5} + 4 x^{4} + 11 x^{3} $ & 105 \\
$ 4 x^{6} + 4 x^{5} + 6 x^{4} + 9 x^{3} $ & 105 \\
$ 4 x^{6} + 4 x^{4} + 15 x^{3} $ & 106 \\
$ 4 x^{6} + 2 x^{5} + 4 x^{4} + 13 x^{3} $ & 107 \\
$ 4 x^{6} + 2 x^{5} + 7 x^{4} + 10 x^{3} $ & 145 \\
$ 4 x^{6} + 4 x^{5} + 5 x^{4} + 10 x^{3} $ & 146 \\
$ 4 x^{6} + 2 x^{5} + 5 x^{4} + 12 x^{3} $ & 174 \\
$ 4 x^{6} + 2 x^{5} + 6 x^{4} + 11 x^{3} $ & 221 \\
$ 4 x^{6} + x^{5} + 8 x^{4} + 10 x^{3} $ & 224 \\
$ 4 x^{6} + x^{5} + 4 x^{4} + 14 x^{3} $ & 342 \\
$ 4 x^{6} + 9 x^{4} + 10 x^{3} $ & 631 \\
$ 4 x^{6} + x^{5} + 5 x^{4} + 13 x^{3} $ & 637 \\
$ 4 x^{6} + x^{5} + 7 x^{4} + 11 x^{3} $ & 700 \\
$ 4 x^{6} + x^{5} + 6 x^{4} + 12 x^{3} $ & 767 \\
$ 4 x^{6} + 5 x^{4} + 14 x^{3} $ & 913 \\
$ 4 x^{6} + 8 x^{4} + 11 x^{3} $ & 2060 \\
$ 4 x^{6} + 6 x^{4} + 13 x^{3} $ & 2121 \\
$ 4 x^{6} + 7 x^{4} + 12 x^{3} $ & 2843 
\end{tabular}

  }
\[  
  x^{\sum_{\ell=1}^{23}\rank(A_\ell)} +
   x^{\sum_{\ell=1}^{23}\rank(B_\ell)} +
   x^{\sum_{\ell=1}^{23}\rank(C_\ell)}
\]
{\footnotesize   
   
\begin{tabular}{r|l}
$ 2 x^{28} + x^{27} $ & 1 \\
$ x^{34} + 2 x^{31} $ & 1 \\
$ x^{33} + x^{32} + x^{30} $ & 1 \\
$ x^{33} + x^{32} + x^{29} $ & 1 \\
$ 2 x^{32} + x^{28} $ & 1 \\
$ x^{34} + 2 x^{32} $ & 3 \\
$ x^{32} + 2 x^{30} $ & 3 \\
$ x^{34} + 2 x^{30} $ & 3 \\
$ x^{33} + x^{31} + x^{29} $ & 4 \\
$ 2 x^{33} + x^{32} $ & 5 \\
$ x^{34} + 2 x^{29} $ & 5 \\
$ x^{34} + x^{30} + x^{29} $ & 6 \\
$ x^{33} + 2 x^{30} $ & 6 \\
$ 3 x^{28} $ & 7 \\
$ 3 x^{29} $ & 7 \\
$ 3 x^{30} $ & 8 \\
$ 2 x^{31} + x^{27} $ & 9 \\
$ x^{29} + 2 x^{28} $ & 9 \\
$ x^{33} + x^{31} + x^{30} $ & 14 
\end{tabular}\hfill\begin{tabular}{r|l}
$ 2 x^{32} + x^{29} $ & 15 \\
$ 2 x^{33} + x^{31} $ & 16 \\
$ x^{30} + x^{28} + x^{27} $ & 16 \\
$ x^{31} + x^{28} + x^{27} $ & 21 \\
$ 2 x^{29} + x^{28} $ & 24 \\
$ x^{33} + x^{30} + x^{29} $ & 24 \\
$ 2 x^{32} + x^{30} $ & 26 \\
$ x^{33} + 2 x^{32} $ & 31 \\
$ x^{33} + x^{30} + x^{28} $ & 37 \\
$ x^{33} + x^{32} + x^{31} $ & 41 \\
$ 2 x^{32} + x^{27} $ & 42 \\
$ x^{33} + 2 x^{31} $ & 45 \\
$ x^{31} + 2 x^{30} $ & 46 \\
$ 2 x^{30} + x^{29} $ & 49 \\
$ x^{29} + x^{28} + x^{27} $ & 51 \\
$ 3 x^{32} $ & 51 \\
$ x^{32} + x^{31} + x^{27} $ & 55 \\
$ x^{32} + x^{31} + x^{28} $ & 56 \\
$ x^{30} + 2 x^{28} $ & 62 
\end{tabular}\hfill\begin{tabular}{r|l}
$ 2 x^{30} + x^{28} $ & 63 \\
$ 2 x^{31} + x^{28} $ & 66 \\
$ x^{30} + 2 x^{29} $ & 70 \\
$ 2 x^{30} + x^{27} $ & 71 \\
$ 3 x^{31} $ & 77 \\
$ 2 x^{29} + x^{27} $ & 85 \\
$ x^{32} + x^{31} + x^{29} $ & 87 \\
$ 2 x^{31} + x^{29} $ & 88 \\
$ x^{33} + x^{29} + x^{27} $ & 96 \\
$ 2 x^{31} + x^{30} $ & 97 \\
$ 2 x^{32} + x^{31} $ & 110 \\
$ x^{32} + x^{31} + x^{30} $ & 113 \\
$ x^{31} + x^{29} + x^{27} $ & 128 \\
$ x^{30} + x^{29} + x^{27} $ & 133 \\
$ x^{31} + 2 x^{28} $ & 133 \\
$ x^{33} + 2 x^{29} $ & 134 \\
$ x^{32} + x^{30} + x^{29} $ & 135 \\
$ x^{30} + x^{29} + x^{28} $ & 137 \\
$ x^{31} + x^{30} + x^{27} $ & 138
\end{tabular}\hfill\begin{tabular}{r|l}
$ x^{32} + 2 x^{31} $ & 186 \\
$ x^{31} + x^{30} + x^{28} $ & 279 \\
$ x^{32} + x^{30} + x^{27} $ & 287 \\
$ x^{31} + x^{30} + x^{29} $ & 291 \\
$ x^{32} + x^{30} + x^{28} $ & 301 \\
$ x^{31} + 2 x^{29} $ & 353 \\
$ x^{33} + x^{29} + x^{28} $ & 364 \\
$ x^{33} + 2 x^{28} $ & 392 \\
$ x^{32} + 2 x^{29} $ & 425 \\
$ x^{33} + 2 x^{27} $ & 485 \\
$ x^{31} + x^{29} + x^{28} $ & 580 \\
$ x^{32} + 2 x^{27} $ & 674 \\
$ x^{32} + x^{29} + x^{27} $ & 704 \\
$ x^{33} + x^{28} + x^{27} $ & 738 \\
$ x^{32} + 2 x^{28} $ & 1230 \\
$ x^{32} + x^{29} + x^{28} $ & 1470 \\
$ x^{32} + x^{28} + x^{27} $ & 1510 \\
\vphantom{$ x^{32} + x^{28} + x^{27} $} & \vphantom{1510} \\
\vphantom{$ x^{32} + x^{28} + x^{27} $} & \vphantom{1510}
\end{tabular}

}

\par\medskip\noindent
 $\blacksquare$\quad A multiplication scheme for the coefficient ring $\set Z_2$ that cannot be extended to a scheme for $\set Z$
 by replacing some of the $1$'s by $-1$'s. It may still be possible to find a scheme with coefficients in $\set Z$ which reduces
 to this scheme modulo~2, but any such scheme must have at least one coefficient with absolute value $\geq2$. 

 \def\mymat#1{\begin{pmatrix}#1\end{pmatrix}}
 \def\row#1#2#3{
   \hbox to.125\hsize{\hss$\mymat{#1}$\hss}$\otimes$\hbox to.125\hsize{\hss$\mymat{#2}$\hss}$\otimes$\hbox to.125\hsize{\hss$\mymat{#3}$\hss}}

{\par\centering\leavevmode 
\hphantom{+}\row{ 0 & 0 & 0 \\ 0 & 0 & 0 \\ 1 & 0 & 0 }{ 0 & 1 & 0 \\ 0 & 1 & 0 \\ 0 & 1 & 0 }{ 1 & 0 & 0 \\ 0 & 1 & 1 \\ 0 & 1 & 1 }
+\row{ 0 & 0 & 0 \\ 0 & 0 & 0 \\ 0 & 0 & 1 }{ 0 & 0 & 0 \\ 1 & 0 & 0 \\ 1 & 0 & 0 }{ 1 & 1 & 1 \\ 0 & 0 & 0 \\ 0 & 0 & 0 }\\
+\row{ 0 & 0 & 0 \\ 0 & 0 & 0 \\ 0 & 1 & 1 }{ 0 & 0 & 0 \\ 1 & 0 & 0 \\ 0 & 0 & 0 }{ 0 & 0 & 1 \\ 0 & 0 & 1 \\ 0 & 0 & 1 }
+\row{ 0 & 0 & 0 \\ 1 & 0 & 0 \\ 0 & 0 & 0 }{ 1 & 0 & 0 \\ 0 & 0 & 0 \\ 0 & 0 & 0 }{ 1 & 1 & 1 \\ 0 & 0 & 0 \\ 0 & 0 & 0 }\\
+\row{ 0 & 0 & 0 \\ 1 & 0 & 0 \\ 1 & 0 & 0 }{ 1 & 0 & 0 \\ 0 & 0 & 0 \\ 0 & 0 & 0 }{ 1 & 0 & 1 \\ 0 & 0 & 0 \\ 0 & 0 & 0 }
+\row{ 0 & 0 & 0 \\ 1 & 0 & 0 \\ 1 & 0 & 0 }{ 0 & 0 & 1 \\ 0 & 0 & 0 \\ 0 & 0 & 0 }{ 0 & 0 & 0 \\ 0 & 1 & 0 \\ 0 & 1 & 0 }\\
+\row{ 0 & 0 & 0 \\ 0 & 1 & 0 \\ 0 & 0 & 0 }{ 0 & 1 & 1 \\ 0 & 1 & 1 \\ 0 & 1 & 1 }{ 0 & 0 & 0 \\ 0 & 0 & 0 \\ 0 & 1 & 1 }
+\row{ 0 & 0 & 0 \\ 0 & 1 & 0 \\ 1 & 0 & 0 }{ 0 & 1 & 1 \\ 0 & 1 & 0 \\ 0 & 1 & 0 }{ 0 & 0 & 0 \\ 0 & 1 & 0 \\ 0 & 1 & 1 }\\
+\row{ 0 & 0 & 0 \\ 0 & 1 & 0 \\ 0 & 1 & 0 }{ 0 & 0 & 0 \\ 1 & 0 & 1 \\ 0 & 0 & 0 }{ 0 & 0 & 0 \\ 0 & 0 & 0 \\ 0 & 0 & 1 }
+\row{ 0 & 0 & 0 \\ 0 & 1 & 0 \\ 0 & 0 & 1 }{ 0 & 0 & 0 \\ 1 & 0 & 0 \\ 0 & 0 & 1 }{ 0 & 1 & 0 \\ 0 & 0 & 0 \\ 0 & 0 & 1 }\\
+\row{ 0 & 0 & 0 \\ 0 & 0 & 1 \\ 0 & 0 & 1 }{ 0 & 0 & 0 \\ 0 & 0 & 0 \\ 1 & 0 & 1 }{ 0 & 1 & 0 \\ 0 & 0 & 0 \\ 0 & 0 & 0 }
+\row{ 0 & 0 & 0 \\ 0 & 1 & 1 \\ 0 & 0 & 0 }{ 0 & 0 & 0 \\ 0 & 0 & 0 \\ 0 & 0 & 1 }{ 0 & 1 & 0 \\ 0 & 0 & 0 \\ 1 & 1 & 0 }\\
+\row{ 1 & 0 & 0 \\ 0 & 0 & 0 \\ 0 & 0 & 0 }{ 0 & 1 & 0 \\ 0 & 1 & 0 \\ 0 & 0 & 0 }{ 1 & 0 & 0 \\ 1 & 0 & 0 \\ 1 & 0 & 0 }
+\row{ 1 & 0 & 0 \\ 0 & 0 & 0 \\ 1 & 0 & 0 }{ 1 & 1 & 0 \\ 1 & 1 & 0 \\ 0 & 0 & 0 }{ 1 & 0 & 0 \\ 0 & 0 & 0 \\ 0 & 0 & 0 }\\
+\row{ 0 & 1 & 0 \\ 0 & 0 & 0 \\ 0 & 0 & 0 }{ 0 & 0 & 0 \\ 0 & 1 & 1 \\ 0 & 1 & 1 }{ 0 & 0 & 0 \\ 0 & 0 & 0 \\ 1 & 0 & 0 }
+\row{ 1 & 1 & 0 \\ 0 & 0 & 0 \\ 0 & 0 & 0 }{ 0 & 0 & 0 \\ 0 & 1 & 0 \\ 0 & 1 & 0 }{ 0 & 0 & 0 \\ 1 & 0 & 1 \\ 1 & 0 & 0 }\\
+\row{ 1 & 1 & 0 \\ 0 & 0 & 0 \\ 1 & 1 & 0 }{ 0 & 0 & 0 \\ 1 & 1 & 0 \\ 0 & 0 & 0 }{ 0 & 0 & 0 \\ 0 & 0 & 1 \\ 0 & 0 & 0 }
+\row{ 1 & 1 & 0 \\ 0 & 0 & 0 \\ 1 & 0 & 1 }{ 0 & 0 & 0 \\ 1 & 0 & 0 \\ 0 & 1 & 0 }{ 1 & 0 & 0 \\ 0 & 0 & 1 \\ 0 & 0 & 0 }\\
+\row{ 0 & 0 & 1 \\ 0 & 0 & 0 \\ 0 & 0 & 1 }{ 0 & 0 & 0 \\ 0 & 0 & 0 \\ 1 & 1 & 0 }{ 1 & 0 & 0 \\ 0 & 0 & 0 \\ 0 & 0 & 0 }
+\row{ 0 & 1 & 1 \\ 0 & 1 & 1 \\ 0 & 0 & 0 }{ 0 & 1 & 1 \\ 0 & 0 & 0 \\ 0 & 1 & 1 }{ 0 & 0 & 0 \\ 0 & 0 & 0 \\ 1 & 0 & 0 }\\
+\row{ 1 & 1 & 1 \\ 0 & 0 & 0 \\ 0 & 0 & 0 }{ 0 & 0 & 0 \\ 0 & 0 & 0 \\ 0 & 1 & 0 }{ 1 & 0 & 0 \\ 1 & 1 & 0 \\ 0 & 0 & 0 }
+\row{ 1 & 1 & 1 \\ 1 & 0 & 1 \\ 0 & 0 & 0 }{ 0 & 1 & 1 \\ 0 & 0 & 0 \\ 0 & 0 & 0 }{ 0 & 0 & 0 \\ 0 & 1 & 0 \\ 0 & 0 & 0 }\\
+\row{ 1 & 1 & 1 \\ 0 & 1 & 1 \\ 0 & 0 & 0 }{ 0 & 1 & 1 \\ 0 & 0 & 0 \\ 0 & 1 & 0 }{ 0 & 0 & 0 \\ 0 & 1 & 0 \\ 1 & 0 & 0 }
\hphantom{+\row{ 1 & 1 & 1 \\ 0 & 1 & 1 \\ 0 & 0 & 0 }{ 0 & 1 & 1 \\ 0 & 0 & 0 \\ 0 & 1 & 0 }{ 0 & 0 & 0 \\ 0 & 1 & 0 \\ 1 & 0 & 0 }}
\par}

\par\medskip\noindent
 $\blacksquare$\quad
 A general multiplication scheme with 17 parameters.
 The parameters are $x_1,\dots,x_{17}$, and we use the following shortcuts:
 \begin{alignat*}3
   x_{i,j} &= x_ix_j + 1 &\qquad p_{1} &= x_9 x_{6,8}+x_6\\
   p_{2} &= x_5 x_{1,4}+x_1 &\qquad p_{3} &= x_{16} x_{3,15}+x_3\\
   p_{4} &= x_{13} x_{11,12}+x_{11} &\qquad p_{5} &= x_2 x_{10} x_{1,4}-x_{10} x_{17} x_{1,4}+x_2 x_4\\
   p_{6} &= \rlap{$x_2 x_5 x_{10} x_{1,4}-x_5 x_{17} x_{10} x_{1,4}+x_2 x_{4,5}+x_1 x_2 x_{10}-x_1 x_{17} x_{10}$}
 \end{alignat*}
 
 \def\mymat#1{\begin{pmatrix}#1\end{pmatrix}}
 \def\row#1#2#3{
   \hbox to.225\hsize{\hss$\mymat{#1}$\hss}$\otimes$\hbox to.325\hsize{\hss$\mymat{#2}$\hss}$\otimes$\hbox to.3\hsize{\hss$\mymat{#3}$\hss}}

{\par\centering\leavevmode  
\hphantom+\row{ 1 & -1 & -1 \\ 1 & 0 & 0 \\ 0 & 0 & 0 }{ 0 & 1 & 0 \\ 0 & 0 & 0 \\ 0 & 0 & 0 }{ -1 & 0 & 0 \\ 1 & 0 & 0 \\ 1 & 0 & 0 }\\
+\row{ 1 & -1 & 0 \\ 0 & 0 & 0 \\ 1 & 0 & 0 }{ 1 & 1 & 0 \\ 0 & 0 & 0 \\ 0 & 0 & 0 }{ 1 & 0 & 0 \\ 0 & 0 & 0 \\ 0 & 0 & 0 }\\
+\row{ x_{4,5} & 0 & p_2 \\ x_{4,5} & 0 & p_2 \\ p_6 & 0 & p_2 x_{17} }{ 0 & -x_{1,4} & x_{1,4} \\ 0 & 0 & 0 \\ 0 & x_4 & -x_4 }{ 0 & 0 & 0 \\ 0 & 0 & 0 \\ 1 & 0 & 0 }\\
+\row{ 0 & 1 & 1 \\ 0 & 0 & 0 \\ 1 & 0 & 0 }{ 0 & 1 & 0 \\ 0 & 0 & 0 \\ -1 & 0 & 0 }{ -1 & 0 & 0 \\ 0 & 0 & 1 \\ 0 & 0 & 1 }\\
+\row{ 0 & 1 & 1 \\ 0 & 0 & 0 \\ 0 & 0 & 1 }{ 0 & 0 & 0 \\ 0 & 0 & 0 \\ 1 & 1 & 0 }{ 0 & 0 & 0 \\ 0 & 0 & 1 \\ 0 & 0 & 1 }\\
+\row{ 0 & 1 & 1 \\ 0 & 0 & 0 \\ 0 & 0 & 0 }{ 0 & 1 & 0 \\ 0 & 0 & 0 \\ 0 & 1 & 0 }{ 0 & 0 & 0 \\ 1 & 0 & -1 \\ 1 & 0 & -1 }\\
+\row{ 0 & 1 & 0 \\ 1 & 0 & 0 \\ 0 & 0 & 0 }{ 1 & 1 & 0 \\ 1 & 0 & 0 \\ -1 & 0 & 0 }{ 1 & 0 & 0 \\ -1 & 1 & 0 \\ 0 & 0 & 0 }\\
+\row{ 0 & x_{15,16} & 0 \\ 0 & x_{15,16} & 0 \\ 0 & p_3 & 0 }{ 0 & 0 & 0 \\ 0 & -1 & 1 \\ 0 & 0 & 0 }{ 0 & x_{14} x_{15} & x_{14} x_{15} \\ 0 & -x_{3,15} & x_{15} \\ 0 & 0 & 0 }\\
+\row{ 0 & 1 & 0 \\ 0 & 0 & -1 \\ 0 & 0 & 0 }{ 0 & 0 & 0 \\ 0 & 0 & 1 \\ 0 & -1 & 0 }{ 0 & 0 & 0 \\ 0 & 1 & 0 \\ 1 & 0 & 0 }\\
+\row{ 0 & 1 & 0 \\ 0 & 0 & 0 \\ 0 & 0 & 0 }{ 1 & 1 & 0 \\ 1 & 1 & 0 \\ -1 & -1 & 0 }{ 0 & 0 & 0 \\ 1 & -1 & 0 \\ 0 & 0 & 0 }\\
+\row{ x_4 & 0 & x_{1,4} \\ x_4 & 0 & x_{1,4} \\ p_5 & 0 & x_{17} x_{1,4} }{ 0 & p_2 & -p_2 \\ 0 & 0 & 0 \\ 0 & -x_{4,5} & x_{4,5} }{ 0 & 0 & 0 \\ 0 & 0 & 0 \\ 1 & 0 & 0 }\\
+\row{ 0 & 0 & 0 \\ 1 & -1 & 0 \\ 0 & 1 & 0 }{ 0 & 0 & 0 \\ -1 & 0 & 0 \\ 1 & 0 & 0 }{ 0 & 1 & 0 \\ 0 & 0 & 0 \\ 0 & 0 & 0 }\\
+\row{ 0 & 0 & 0 \\ p_1 & 0 & x_{8,9} \\ 0 & 0 & 0 }{ x_7 x_8 & 0 & x_8 \\ x_7 x_8 & 0 & x_{6,8} \\ -x_7 x_8 & 0 & -x_{6,8} }{ 0 & 0 & 0 \\ 0 & 0 & 0 \\ 1 & -1 & 0 }\\
+\row{ 0 & 0 & 0 \\ 1 & 0 & 0 \\ 0 & 0 & 0 }{ 1 & 0 & 0 \\ 1 & 0 & 0 \\ -1 & 0 & 0 }{ -1 & 1 & 0 \\ 1 & -1 & 0 \\ x_7 & -x_7 & 0 }\\
+\row{ 0 & 0 & 0 \\ x_{6,8} & 0 & x_8 \\ 0 & 0 & 0 }{ x_7 x_{8,9} & 0 & x_{8,9} \\ x_7 x_{8,9} & 0 & p_1 \\ -x_7 x_{8,9} & 0 & -p_1 }{ 0 & 0 & 0 \\ 0 & 0 & 0 \\ -1 & 1 & 0 }\\
+\row{ 0 & 0 & 0 \\ 0 & 1 & 1 \\ 1 & -1 & -1 }{ 0 & 0 & 0 \\ 0 & 0 & 0 \\ 1 & 0 & 0 }{ 0 & 0 & -1 \\ 0 & 0 & 1 \\ 0 & 0 & 1 }\\
+\row{ 0 & 0 & 0 \\ 0 & 1 & 1 \\ 0 & -1 & 0 }{ 0 & 0 & 0 \\ 0 & 0 & 1 \\ 1 & 0 & 0 }{ 0 & 1 & 1 \\ 0 & 0 & -1 \\ 0 & 0 & -1 }\\
+\row{ 0 & 0 & 0 \\ 0 & 1 & 1 \\ 0 & 0 & 0 }{ 0 & 0 & 0 \\ 0 & 0 & 1 \\ 0 & 0 & 0 }{ 0 & -1 & -1 \\ 0 & 1 & 1 \\ 0 & 1 & 1 }\\
+\row{ 0 & 0 & 0 \\ 0 & 0 & 0 \\ 1 & 0 & 0 }{ p_4 & x_{12,13} & -x_{12,13} \\ 0 & 0 & 0 \\ p_4 & x_{10} x_{12,13} & -x_{10} x_{12,13} }{ x_{12} & 0 & -x_{12} \\ 0 & 0 & 0 \\ x_2 x_{11,12} & 0 & -x_{11,12} }\\
+\row{ 0 & 0 & 0 \\ 0 & 0 & 0 \\ -1 & 0 & 0 }{ x_{11,12} & x_{12} & -x_{12} \\ 0 & 0 & 0 \\ x_{11,12} & x_{10} x_{12} & -x_{10} x_{12} }{ x_{12,13} & 0 & -x_{12,13} \\ 0 & 0 & 0 \\ p_4 x_2 & 0 & -p_4 }\\
+\row{ 0 & 0 & 0 \\ 0 & 0 & 0 \\ 0 & 1 & 0 }{ 0 & 0 & 0 \\ 1 & -x_{14} & x_{14}+1 \\ 0 & 0 & 0 }{ 0 & 1 & 1 \\ 0 & 0 & 0 \\ 0 & 0 & 0 }\\
+\row{ 0 & x_{15} & 0 \\ 0 & x_{15} & 0 \\ 0 & x_{3,15} & 0 }{ 0 & 0 & 0 \\ 0 & 1 & -1 \\ 0 & 0 & 0 }{ 0 & x_{14} x_{15,16} & x_{14} x_{15,16} \\ 0 & -p_3 & x_{15,16} \\ 0 & 0 & 0 }\\
+\row{ 0 & 0 & 0 \\ 0 & 0 & 0 \\ -x_{10} & 0 & 1 }{ 0 & 0 & 0 \\ 0 & 0 & 0 \\ 0 & 1 & -1 }{ 0 & 0 & 0 \\ 0 & 0 & 0 \\ x_{17} & 0 & -1 }
\par}


\end{document}